\newcommand{\sbf}[1]{\mbox{{\scriptsize$\bf{#1}$}}}
\def\Bf#1{\mbox{\boldmath $#1$}}
\def\comment#1{}
\def\lfrac#1#2{#1/#2}

\documentstyle[pra,epsf,aps]{revtex}

\def\cm#1{}
\def\sbf#1{\mbox{\scriptsize{\bf #1}}}

 \def\lfrac#1#2{{{{#1}/{#2}}}}

\begin{document}
\title{{Solution of Coulomb Path Integral in Momentum Space
}}
\author{Hagen Kleinert%
 \thanks{Email: kleinert@physik.fu-berlin.de  ~~ URL:
http://www.physik.fu-berlin.de/\~{}kleinert \hfil
}}
\address{Institut f\"ur Theoretische Physik,\\
Freie Universit\"at Berlin, Arnimallee 14,
14195 Berlin, Germany}
\maketitle
\begin{abstract}
The path integral for a point particle in a Coulomb
potential is solved in momentum space.
The solution permits us to
give
for the first time
a negative  answer to
an old  question
of quantum mechanics in curved spaces
raised in 1957 by DeWitt, whether
the Hamiltonian of a particle in a curved space
contains an additional
term proportional
to the curvature scalar $R$. We show that this
would cause experimentally wrong
{\em level spacings\/} in the hydrogen atom.
Our solution also gives a first experimental confirmation
of the correctness of the measure of integration
in path integrals in curved space
implied by a recently discovered nonholonomic mapping principle.
\end{abstract}

{}~\\
{\bf 1.} One should think that
by now everything interesting is known about the
path integral of the Coulomb problem describing
the physics of the hydrogen atom.
There exists a comprehensive textbook \cite{PI} in which this
subject is treated at great length.
However, the existing solution
applies only to the fixed-energy amplitude in
position space.
The momentum space problem has so far remained untackled,
and the purpose of this note is to fill this gap.

Apart from our desire to complete the path integral description of
the simplest physical object of atomic physics,
the present note is motivated
by another long-standing open problem
in the quantum mechanics of curved spaces, first raised by Bryce DeWitt
in 1957 \cite{DW}:
Is the Hamilton operator for a particle
in curved space
obtained by merely replacing the euclidean Laplace operator
in the kinetic energy by
the Laplace-Beltrami operator $ \Delta $,
or must we add a term proportional
to $\hbar ^2R$, as suggested by various path integral
formulations of the problem in the literature \cite{DW,cheng,var}?
Only
experiment can decide what is right, but
up to now no physical system has been contrived
where the presence of an extra $R$-term could be detectable.
All experimentally accessible systems
in curved space
have either
a very small
$R$ caused by gravitation, whose
detection is presently impossible, or a
 constant $R$ which does not change level spacings,
an example for the latter being
the spinning symmetric and asymmetric top \cite{PI}.
Surprisingly, the solution developed in this note
supplies an answer to this problem by
forbidding an extra $R$-term, which although being constant
 would change the level spacings in the hydrogen atom.

{}~\\
{\bf 2.} Starting point for our treatment is
the path integral formulation
for the matrix elements in momentum space
fo the resolvent operator $\hat R\equiv i/(E-\hat H)$
with the Coulomb Hamilton operator $\hat H=\hat {\bf p}^2/2- \alpha /r$.
We use natural units with $\hbar =M=1$,
so that masses, lengths, times, and energies will have the units of
$M,~\hbar ^2/M \alpha ,~\hbar ^3/M \alpha ^2$, and
$M \alpha ^2/\hbar ^2\approx27.21$ eV, respectively.
The resolvent can be reexpressed
as
\begin{equation}
 \hat R=\frac{i}{\hat f(E-\hat H) }\hat f
\label{@}\end{equation}
where $f$ is an arbitrary function of space, momentum, and some parameter
$s$ \cite{sm}.
Using standard techniques \cite{PI},
the matrix elements of the resolvent are represented by
the following canonical euclidean path integral:
\begin{eqnarray}
\!\!\!\!\!\!\!\!({\bf p}_b|{\bf p}_a)_E^f&=&\int _0^\infty dS
\int \frac{{\cal D}^3p}{(2\pi)^3}\int {\cal D}^3x
\exp\left\{- \int_0^Sds \left[i \dot {\bf p}\cdot{\bf x}+f
\left( \frac{{\bf p}^2}{2}-E\right)-f {\alpha}
\right]\right\} f(0) .
\label{@amp1}\end{eqnarray}
The dot denotes differentiation with respect to $s$.
The left-hand side carries
a superscript $f$ to remind us of
the presence of $f$
on the right-hand side, although
the amplitude
does not really depend on $f$.
This freedom of choice
may be viewed as a
gauge invariance \cite{fuji}
of (\ref{@amp1}) under $f\rightarrow f'$. It permits us to
subject (\ref{@amp1})
to an additional  path integration
over $f$, as long as a gauge fixing functional
$\Phi[f]$ ensures that only a specific ``gauge"
contributes. Thus we shall calculate the amplitude as
a path integral
\begin{equation}
({\bf p}_b|{\bf p}_a)_E=\int{\cal D} f\,\Phi[f]
{}~({\bf p}_b|{\bf p}_a)_E^f.
\label{@3}\end{equation}
The only condition on $\Phi[f]$ is that $\int{\cal D} f\,\Phi[f]=1$.
The choice which will lead to the desired solution of the path integral is
\begin{equation}
\Phi[f]= \prod_{s}\frac{1}{r}\exp\left\{ -\frac{1}{2r^2}\left[f-{\bf
x}^2\left(\frac{{\bf p}^2}{2}-E\right)\right]^2\right\} .
\label{@}\end{equation}
With this, the total euclidean action in the path integral
(\ref{@3}) is
\begin{eqnarray}
{\cal A}_{\rm e}[{\bf p},{\bf x},h]=\int_0^Sds \left[i \dot {\bf p}\cdot{\bf
x}+\frac{1}{2}
{\bf x}^2\left( \frac{{\bf p}^2}{2}-E\right)^2
+\frac{1}{2r^2}f^2 -\frac{1}{r}f{\alpha}
\right]
{}.
\label{@ea1}\end{eqnarray}
The path integrals over $f$ and  ${\bf x}$ in (\ref{@amp1})
are Gaussian and can be done, in this order,
yielding a new
euclidean action
\begin{eqnarray}
{\cal A}_{\rm e}[{\bf p}]=\frac{1}{2}\int_0^Sds \left[\frac{4\dot{\bf
p}^2}{\left({\bf p}^2+p_E^2\right)^2}
- \alpha ^2
\right],
\label{@ea2}\end{eqnarray}
where we have introduced $p_E= \sqrt{-2E}$,
assuming $E$ to be negative. The positive regime can always be
obtained  by analytic continuation.
Now, a stereographic projection
\begin{eqnarray}
\Bf{\pi}\equiv \frac{2p_E{\bf p}}{{\bf p}^2+p_E^2},~~~~
\pi_4\equiv \frac{{\bf p}^2-p_E^2}{{\bf p}^2+p_E^2}
\label{@pis}\end{eqnarray}
transforms (\ref{@ea2})
to the form
\begin{eqnarray}
{\cal A}_{\rm e}[\vec\pi]=\frac{1}{2}\int_0^Sds \left(
\frac{1}{p_E^2}
\dot
{\vec{\pi}}^2
- \alpha ^2
\right),
\label{@ea2}\end{eqnarray}
where $\vec\pi$ denotes the four-dimensional
unit vectors $({\Bf \pi},\pi_4)$.
 This describes a point particle of pseudomass $\mu=1/p_E^2$
moving on a four-dimensional
unit sphere.
The pseudotime evolution amplitude
of this system is
\begin{equation}
({\vec \pi}_bS|{\vec \pi}_a0)
=e^{-Sp_E^2}\int \frac{{\cal D}^4\pi}{(2\pi)^{3/2}p_E^3}\,
e^{-{\cal A}_{\rm e}[\vec \pi]}.
\label{@sos}\end{equation}
There is an exponential prefactor
arising
from the transformation of the functional measure
in (\ref{@amp1}) to the unit sphere.
Let us see how this comes about.
When integrating out the spatial fluctuations in going from
(\ref{@ea1}) to (\ref{@ea2}),
the canonical measure in each time slice
$
\lfrac{{{ d}^3p}\, { d}^3x}{(2\pi)^3}$
becomes
$\lfrac{{d}^3p\,8}{(2\pi)^{3/2}({\bf p}^2+p_E^2)}$.
{}From the stereographic projection
(\ref{@pis}) we see that this is equal to
$\lfrac{{d}^4\vec \pi}{(2\pi)^{3/2}p_E^3}$,
where ${d}^4\vec \pi$ denotes the product of
integrals over the solid angle on the surface
of the unit sphere in four dimensions,
with the integral $\int {d}^4\vec \pi$
yielding the total surface $2\pi^2$.
{}From Chapter 10 in the textbook \cite{PI}
we know that in a curved space,
the time sliced measure of path integration
is given by the product of invariant integrals
$\int dq \sqrt{g(q)}$ in each time slice,
multiplied by an effective
action
contribution
$\exp(-{\cal A}_{\rm eff})=
\exp({\int ds\bar R/6\mu})$,
where $\bar R$ is the scalar curvature. For a sphere of radius $r$
in $D$ dimensions,
$\bar R=(D-1)(D-2)/r^2$, implying here
$
\exp(-{\cal A}_{\rm eff})
=\exp(\int ds\, 1/\mu)
=\exp(\int ds \,p_E^2)$.
Thus, when transforming the time-sliced measure
in the original path integral
(\ref{@3})  to
the time-sliced measure
on the sphere in (\ref{@sos})
which contains the effective action,
the exponent is modified accordingly.

A complete set of orthonormal hyperspherical functions
on this sphere may be denoted by $Y_{nlm}(\vec \pi)$, where $n,l,m$ are the
quantum numbers of the
hydrogen atom with the well-known ranges
$(n=1,2,3,\dots,~l=0,\dots,n-1,~m=-l,\dots,l)$.
They can be expressed in terms of the
three-dimensional representation
$D^j_{m_1m_2}(u)$ of the SU(2) matrices
$u=\vec \pi\vec \sigma$ with the Pauli matrices $\vec  \sigma \equiv (1,
 \sigma ^1,  \sigma  ^2, \sigma ^3)$
as
\begin{eqnarray}
Y_{2j+1,l,m}(\vec \pi)= \sqrt{\frac{2j+1}{2\pi^2}}\sum _{m_1,m_2=-j,\dots,j}
(j,m_1;j,m_2|l,m)\, D^j_{m_1m_2}(u).
\label{@}\end{eqnarray}
The orthonormality and completeness relations are
\begin{equation}
\int d\vec\pi\,Y_{n'l'm'}^*(\vec \pi)Y_{nlm}(\vec \pi)=  \delta _{nn'}\delta
_{ll'} \delta _{mm'},~~~~~
\sum_{n,l,m}Y_{nlm}(\vec \pi')Y_{nlm}(\vec \pi)=
\delta^{(4)} (\vec \pi'-\vec\pi).
\label{@}\end{equation}
where the $ \delta $-function satisfies
 $\int d\vec\pi\,\delta^{(4)} (\vec \pi'-\vec\pi)=1$.
When restricting the complete sum to  $l$ and $m$ only
we obtain the four-dimensional analog of the Legendre polynomial:
\begin{equation}
\sum_{l,m}Y_{nlm}(\vec \pi')Y_{nlm}(\vec \pi)= \frac{n^2}{2\pi^2}
P_n(\cos\vartheta)
,~~~~~~
 P_n(\cos\vartheta)=\frac{\sin n\vartheta}{n\sin\vartheta} ,
\label{@}\end{equation}
where $\vartheta$ is the angle
between the four-vectors
$\vec\pi_b$ and $\vec\pi_a$:
\begin{equation}
\cos  \vartheta =
\vec\pi_b\vec\pi_a=
\frac{({\bf p}_b^2-p_E^2)({\bf p}_a^2-p_E^2)+4p_E^2{\bf p}_b{\bf p}_a}{({\bf
p}_b^2+p_E^2)({\bf p}_a^2+p_E^2)}
\label{@}\end{equation}

The path integral for a particle on the surface of a sphere
was solved
in \cite{PI}. The solution
of (\ref{@sos}) reads
\begin{equation}
({\vec \pi}_bS|{\vec \pi}_a0)
=(2\pi)^{3/2}p_E^3
\sum_{n=1}^\infty \frac{n^2}{2\pi^2}P_n(\cos\vartheta)
\,
\exp\left\{ \left[ - p_E^2 n^2+ \alpha ^2\right] \frac{S}{2}\right\} .
\label{@sol}\end{equation}
For the path integral itself in (\ref{@sos}),
the exponential contains the eigenvalue of the squared
angular-momentum operator $ \hat L^2/2\mu$ which in $D$ dimensions
is $l(l+D-2)/2\mu,~~l=0,1,2,\dots~$. In our system with $D=4$, $l=n-1$,
these eigenvalues are $n^2-1$, leading to an exponential $e^{-p_E^2(n^2-1)S}$.
Together with the
exponential  prefactor
in (\ref{@sos}),
this leads to the
exponential in (\ref{@sol}).
The integral over $S$ in (\ref{@amp1}) with (\ref{@amp1})
can now be done
yielding the
amplitude
at zero fixed pseudoenergy
\begin{eqnarray}
({\vec \pi}_b|{\vec \pi}_a)_0&=& -
(2\pi)^{3/2}p_E^3
\sum_{n=1}^\infty
\frac{n^2}{2\pi^2}P_n(\cos\vartheta)
\,
\frac{2}{2E n^2+ \alpha ^2} .
\label{@amp1}\end{eqnarray}
This has poles displaying the hydrogen spectrum at energies:
\begin{equation}
E_n=-\frac{1}{2n^2},~~~n=1,2,3,\dots~.
\label{@sp}\end{equation}
{}~\\{\bf 3.}~Consider the following generalization of the
final action (\ref{@ea2}):
\begin{eqnarray}
{\cal A}_{\rm e}[{\bf p}]=\frac{1}{2}\int_0^Sds
\left[\frac{1}{h}\frac{4\dot{\bf p}^2}{\left({\bf p}^2+p_E^2\right)^2}
- \alpha ^2h
\right],
\label{@ea2p}\end{eqnarray}
This action is invariant under reparametrizations
$s\rightarrow s' $ if simultaneously
  $h\rightarrow h ds/ds'$.
The path integral with the action (\ref{@ea2}) in the
exponent
may thus be rewritten
as a path integral with
the gauge-invariant action (\ref{@ea2p})
and an additional path integral $\int dh\,\Phi[h]$ with an arbitrary
gauge-fixing functional $\Phi[h]$.
Going back to a real-pseudotime
parameter $s=i\tau $,
the action corresponding to (\ref{@ea2p})
which describes the dynamics of the point particle in the Coulomb potential
reads
\begin{eqnarray}
{\cal A}[{\bf p}]=\frac{1}{2}\int_{\tau _a}^{\tau _b}d\tau
\left[\frac{1}{h}\frac{4\dot{\bf p}^2}{\left({\bf p}^2+p_E^2\right)^2}
+ \alpha ^2h
\right],
\label{@ea2pp}\end{eqnarray}
At the extremum in $h$,
this action reduces to
\begin{eqnarray}
{\cal A}[{\bf p}]=2 \alpha \int_{\tau _a}^{\tau _b}d\tau
 \sqrt{
\frac{\dot{\bf p}^2}{
\left({\bf p}^2+p_E^2\right)^2}}
{}.
\label{@ea2ppp}\end{eqnarray}
This is the manifestly reparametrization invariant
form of an action in a curved space with a metric $g^{\mu \nu }= \delta ^{\mu
\nu }/
\left({\bf p}^2+p_E^2\right)^2$.
In fact, this action coincides with the classical eikonal in momentum space:
\begin{equation}
S({\bf p}_b,{\bf p}_a;E)=-\int_{{\sbf p}_a}^{{\sbf p}_b} d\tau \,\dot{\bf p}
{\bf x}.
\label{eik}\end{equation}
Observing that the central attractive force makes
$\dot{\bf p}$ point in the direction $-{\bf x}$,
and inserting $r= \alpha ({\bf p}^2+p_E^2)/2$,
we find precisely the action (\ref{@ea2ppp}).
In fact, the canonical quantization of
a system with the action
(\ref{@ea2ppp}) a la Dirac
leads directly to a path integral
with action (\ref{@ea2pp}) \cite{PI19}.

The eikonal (\ref{eik}), and thus the action (\ref{@ea2ppp}),
determines the classical orbits via the
first extremal principle of theoretical mechanics
found in
1744
by Maupertius.
{}~\\~\\{\bf 4.}~Since the Coulomb path integral in momentum space
is equivalent to that of a point particle on a sphere,
we can use it to pass an experimental judgement
on the possible presence of an extra $R$-term in the
Hamiltonian operator of the Schr\"odinger equation
in curved space
which could be caused by
various historic choices
of the measure of path integration in the literatue
\cite{DW,cheng,var}.
In the exponent
of (\ref{@sol}), an extra term $c\times\mu \hbar ^2R/2 $
in the Hamilton operator
in addition to the Laplace-Beltrami term
$-\mu \hbar ^2\Delta /2$
would appear as an extra
constant $3c$ added to $n^2$.
The hydrogen spectrum would then have the energies
$E_n=-1/2(n^2+3c)$. The only theoretically
proposed candidates for $c$ are $1/24,~1/12$,~and $1/8$
\cite{DW,cheng,var}.
The resulting strong
distortions of the hydrogen spectrum would
certainly have been noticed
experimentally a long time ago,
apart from the fact that
they would contradict
Schr\"odinger theory in $x$-space
whose
spectrum (\ref{@sp}) as the
 first
 triumph of quantum theory in atomic physics.

On fundamental level, the present discussion
confirms the validity of the
nonholonomic mapping principle \cite{PI,nonh}
which predicted the extra factor
$\exp(-{\cal A}_{\rm eff})=
\exp({\int ds\bar R/6\mu})$
in the measure of the path integral
in curved space, without
which the correct spectrum in curved momentum space would not
have been obtained--the energy would have
had the unphysical form $- \alpha /2(n^2-1)$
with a singularity at $n=1$!

{}~\\
{}~\\
{ACKNOWLEDGMENT}~\\
This work was supported by Deutsche Forschungsgemeinschaft
under contract Kl-256.

\end{document}